%% file: ITSC23.tex
\documentclass[letterpaper,10 pt,conference,dvipsnames]{ieeeconf}
\IEEEoverridecommandlockouts

\let\labelindent\relax

\input{packages}
\input{def}
\input{acronyms}

\renewcommand{\baselinestretch}{0.96}
\theoremstyle{definition}
\newtheorem{assumption}{Assumption}
\newtheorem*{assumption*}{Assumption}
\newtheorem{theorem}{Theorem}

\newtheorem{corollary}[theorem]{Corollary}

\newtheorem{definition}{Definition}
\newtheorem{example}{Example}
\theoremstyle{remark}
\newtheorem*{remark}{Remark}

\Crefname{figure}{Fig.}{Figures}
\makeatletter
    \if@cref@capitalise
        \crefname{subsection}{Section}{Sections}
        \crefname{subsubsection}{Section}{Sections}
        \crefname{assumption}{Assumption}{Assumptions}
        \crefname{problem}{Problem}{Problems}
    \else
        \crefname{subsection}{section}{sections}
        \crefname{subsubsection}{section}{sections}
        \crefname{assumption}{assumption}{assumptions}
        \crefname{problem}{problem}{problems}
    \fi
\makeatother

\usetikzlibrary{positioning,intersections,hobby,patterns,calc,decorations.pathmorphing,decorations.markings, shadows,shapes}
\tikzstyle{block} = [draw, rectangle, minimum height=2em, minimum width=3em,font=\bfseries,rounded corners,thick]
\tikzstyle{block1} = [draw, rectangle, minimum height=1.5em, minimum width=2.5em]
\tikzstyle{blockDyn} = [draw, rectangle, minimum height=2.5em, minimum width=3.5em, align=center, inner sep=10pt, thick, fill=white, copy shadow={draw=black,fill=black,opacity=1,shadow xshift=0.5ex,shadow yshift=-0.5ex}]
\tikzstyle{blockAlg} = [draw, rectangle, minimum height=1.5em, minimum width=2.5em, align=center, inner sep=10pt, thick]
\tikzstyle{sum} = [draw,circle]

\tikzstyle{nodePre} = [circle, draw,inner sep=1pt,node contents={$\preceq$},thick]
\tikzstyle{nodePreEmpty} = [circle, draw,inner sep=1pt,thick]
\tikzstyle{nodePos} = [circle, draw,inner sep=1pt,node contents={$\posceq$},thick]
\tikzstyle{nodeProd} = [rectangle, draw,inner sep=4pt,node contents={$\times$},rounded corners,thick]
\tikzstyle{nodeSum} = [rectangle, draw,inner sep=4pt,node contents={$\mathbf{+}$},rounded corners,thick]

\definecolor{DPgreen}{rgb}{0.0, 0.5, 0.0}
\definecolor{red}{rgb}{0.75, 0.0, 0.0}

\newif\ifmargincomments %
\margincommentstrue

\newif\ifextendedversion %

\ifmargincomments

\else

\fi

\title{
\textbf{Strategic Interactions in Multi-modal Mobility Systems: \\ 
A Game-Theoretic Perspective}
}
\author{Gioele Zardini$^{1}$, Nicolas Lanzetti$^{2}$, Giuseppe Belgioioso$^{2}$, Christian Hartnik$^{2}$,\\
Saverio Bolognani$^{2}$, Florian D\"orfler$^{2}$, Emilio Frazzoli$^{1}$
\thanks{
$^{1}$Institute for Dynamic Systems and Control, ETH Z\"urich, Switzerland, {\tt \{gzardini, efrazzoli\}@ethz.ch}.}
\thanks{
$^{2}$Automatic Control Laboratory, ETH Z\"urich, Switzerland, {\tt \{lnicolas,hartnikc,gbelgioioso,bsaverio,dorfler\}\newline@ethz.ch}.}
\thanks{The first two authors contributed equally to this work.}
\thanks{
This work was supported by the Swiss National Science Foundation under NCCR Automation, grant agreement 51NF40\_180545.}
}

\begin{document}

\begin{textblock*}{\textwidth}(15mm,18mm) %
\bf \textcolor{NavyBlue}{To appear in the Proceedings of the 2023 IEEE 26th International Conference on Intelligent Transportation Systems}
\end{textblock*}

\maketitle
\begin{abstract}
The evolution of existing transportation systems, mainly driven by urbanization and increased availability of mobility options, such as private, profit-maximizing ride-hailing companies, calls for tools to reason about their design and regulation.
To study this complex socio-technical problem, one needs to account for the strategic interactions of the heterogeneous stakeholders involved in the mobility ecosystem and analyze how they influence the system.
In this paper, we focus on the interactions between citizens who compete for the limited resources of a mobility system to complete their desired trip. 
Specifically, we present a game-theoretic framework for multi-modal mobility systems, where citizens, characterized by heterogeneous preferences, have access to various mobility options and seek individually-optimal decisions.
We study the arising game and prove the existence of an equilibrium, which can be efficiently computed via a convex optimization problem.
Through both an analytical and a numerical case study for the classic scenario of Sioux Falls, USA, we illustrate the capabilities of our model and perform sensitivity analyses.
Importantly, we show how to embed our framework into a ``larger'' game among stakeholders of the mobility ecosystem (e.g., municipality, Mobility Service Providers (MSPs), and citizens), effectively giving rise to tools to inform strategic interventions and policy-making in the mobility ecosystem.
\end{abstract}

\input{chapters/introduction}
\input{chapters/model}
\input{chapters/results}

\input{chapters/conclusion}

\bibliographystyle{IEEEtran}
\bibliography{paper}

\appendix
\input{chapters/appendix}

\end{document}

%% file: packages.tex
\usepackage[absolute,overlay]{textpos}
\usepackage{times}
\usepackage{enumitem}

\usepackage[usenames,dvipsnames]{xcolor}

\usepackage{amsmath,amssymb,mathtools,nccmath}

\usepackage{amsthm}
\usepackage{bbold,dsfont,bbm}
\allowdisplaybreaks

\theoremstyle{definition}

\usepackage{graphicx}
\usepackage{tikz,tikz-cd,pgfplots}
\pgfplotsset{compat=1.17}
\usepackage{subfig}

\usepackage[hidelinks]{hyperref}
\usepackage[capitalise,noabbrev]{cleveref}
\usepackage{cite}
\usepackage{todonotes}

\usepackage{multirow}
\usepackage{booktabs}
\usepackage[acronym]{glossaries}

\usepackage{comment}

\usepackage[per-mode=symbol]{siunitx}
\DeclareSIUnit{\eur}{\euro}
\DeclareSIUnit{\usd}{USD}
\DeclareSIUnit{\mph}{mph}
\DeclareSIUnit{\month}{month}
\DeclareSIUnit{\year}{year}
\DeclareSIUnit{\million}{Mil}
\DeclareSIUnit{\mile}{mile}
\DeclareSIUnit{\car}{car}
\DeclareSIUnit{\train}{train}
\DeclareSIUnit{\mmveh}{\text{$\mu$}MV}
\DeclareSIUnit{\nounit}{-}
\usepackage[font=footnotesize]{caption}

%% file: acronyms.tex
\definecolor{lightblue}{rgb}{0.60784,0.76078,0.90196}
\definecolor{darkblue}{rgb}{0.26667,0.44706,0.76863}
\definecolor{lightgreen}{rgb}{0.66275,0.81569,0.55686}
\definecolor{darkgreen}{rgb}{0.43922,0.67843,0.27843}
\definecolor{orange}{rgb}{0.92941,0.49020,0.19216}
\definecolor{yellow}{rgb}{1.00000,0.75294,0.00000}
\definecolor{grey}{rgb}{0.64706,0.64706,0.64706}
\definecolor{purple}{rgb}{0.51373,0.23529,0.04706}
\newacronym{abk:amod}{AMoD}{Autonomous Mobility-on-Demand}
\newacronym{abk:iamod}{\mbox{I-AMoD}}{intermodal \gls{abk:amod}}
\newacronym{abk:av}{\mbox{AV}}{autonomous vehicle}
\newacronym{abk:sv}{\mbox{SV}}{standard vehicle}

\newacronym{abk:br}{BR}{Best Response}
\newacronym{abk:bpr}{BPR}{Bureau of Public Roads}
\newacronym{abk:bev}{BEV}{Battery Electric Vehicle}
\newacronym{abk:hev}{HEV}{Hybrid Electric Vehicle}
\newacronym{abk:ca}{CA}{congestion-aware}
\newacronym{abk:cara}{CARS}{congestion-aware routing scheme}
\newacronym{abk:cpo}{CPO}{complete partial order}
\newacronym{abk:cdp}{CDP}{co-design problem}
\newacronym{abk:cdpi}{CDPI}{co-design problem with implementation}
\newacronym{abk:dp}{DP}{design problem}
\newacronym{abk:dpi}{DPI}{design problem with implementation}
\newacronym{abk:dcpo}{DCPO}{directed complete partial order}
\newacronym{abk:es}{ES}{e-scooter}
\newacronym{abk:ffcs}{FFCS}{free floating car sharing systems}
\newacronym{abk:ghg}{GHG}{greenhouse gas}
\newacronym{abk:icev}{ICEV}{Internal Combustion Engine Vehicle}
\newacronym{abk:kpi}{KPIs}{Key Performance Indicators}
\newacronym{abk:lw}{LW}{Lightweight}
\newacronym{abk:mm}{{$\mu$}M}{micromobility}
\newacronym{abk:mod}{MoD}{Mobility-on-Demand}
\newacronym{abk:msp}{MSP}{Mobility Service Provider}
\newacronym{abk:mcdp}{MCDP}{Monotone Co-Design Problem}
\newacronym{abk:mcfp}{MCFP}{multi-commodity flow problem}
\newacronym[plural=NE,firstplural=Nash Equilibria (NE)]{abk:ne}{NE}{Nash Equilibrium}
\newacronym{abk:nyc}{NYC}{New York City}
\newacronym{abk:poset}{poset}{partially ordered set}
\newacronym{abk:sb}{SB}{shared bike}
\newacronym{abk:spp}{SPP}{shortest path problem}
\newacronym{abk:kdspp}{k-dSPP}{k-disjoint \gls{abk:spp}}
\newacronym{abk:su}{SU}{Sport Utility}
\newacronym{abk:SE}{SE}{Stackelberg Equilibrium}
\newacronym{abk:sdp}{SDP}{short-distance price}
\newacronym{abk:ldp}{LDP}{long-distance price}

\newcommand{\true}[1]{\texttt{T}}

\newcommand{\reals}{\mathbb{R}}
\newcommand{\nonnegreals}{\reals_{\geq 0}}
\newcommand{\availableveh}[2]{C_{#1,\mathrm{a}}^#2}
\newcommand{\displaceveh}[2]{C_{#1,\mathrm{d}}^#2}
\newcommand{\maxrideveh}[3]{C_{#1#2,\mathrm{r}}^#3}

\newcommand{\costride}[3]{c_{#1#2}^#3}

%% file: chapters/introduction.tex
\section{Introduction}
\label{sec:introduction}
In the past years, travel needs from, to, and within urban environments have dramatically increased~\cite{czepkiewicz2018urbanites}, mainly due to urbanization (by 2050, 68\% of the world's population will reside in cities~\cite{un2020}).
This rapidly expanding setting poses several critical questions, and cities need to take important decisions, adapting current transportation systems to welcome larger and changing travel demands.
In particular, cities need to reshape current infrastructure and devise policies to improve the performance of mobility systems, while ensuring accessibility, fairness, and equity~\cite{ranchordas2020smart}.
Furthermore, the impact of policies on private \glspl{abk:msp} such as \gls{abk:mm} operators, ride-hailing companies, and, in the near future, \gls{abk:amod} systems, requires a careful analysis~\cite{zardinilanzettiAR2021}.
Indeed, while offering more choices to travelers, such systems often operate benefiting from public resources (such as public spaces and roads), are profit-oriented, and could lead to disruptive consequences for transportation systems and society at large~\cite{berger2018drivers,rogers2015social}.
Finally, cities are estimated to be responsible for 78\% of the world's energy consumption and for more than 60\% of the global greenhouse emissions~\cite{un2020bis}, posing important sustainability goals for the implemented regulations.

These perspectives highlight the complexity of this socio-technical system and call for tools to inform and drive policymakers of the future.
In this paper, we present a game-theoretic framework to study citizens' behavior in multi-modal mobility systems from a game-theoretic standpoint, and provide both analytical and numerical examples to illustrate its features. 
Additionally, we show how our framework can be integrated into an ``upper-level'' game among stakeholders of the mobility ecosystem, recently proposed in the literature~\cite{zardini2021game}, which can be readily used to reason about policy-making in transportation systems. 

\paragraph{Related Literature}
Our work lies at the interface of game-theoretic modeling of transportation systems and policy-making related to future mobility systems.
Ever since the pioneering work of Pigou~\cite{pigou1924economics}, applications in transportation systems have driven the development of many tools in game theory, such as Wardrop equilibria~\cite{wardrop1952road} and congestion games~\cite{rosenthal1973class}. 
From the study of traffic equilibria~\cite{dafermos1969traffic}, to explaining apparent paradoxes~\cite{braess1968paradoxon}, or to quantifying inefficiencies of selfish routing~\cite{roughgarden2002bad}, game theory has been widely used to model users' behavior in mobility systems~\cite{hollander2006applicability,fisk1984game}.
This line of research, and in particular the theory of non-atomic routing games with capacitated networks~\cite{roughgarden2005selfish,correa2004selfish}, will serve as the basis for our work.

Game theory has also been leveraged for policy-making, for instance for designing pricing strategies for \glspl{abk:msp}~\cite{kuiteing2017network,bimpikis2019spatial,yang2019subsidy,chen2016management}, incentives and tolls to regulate congested networks~\cite{swamy2012effectiveness, paccagnan2021optimal,lazar2020optimal,krichene2017stackelberg,bianco2016game}, analysis of interactions between authorities and \glspl{abk:msp}~\cite{di2019unified,mo2021dynamic,lanzetti2019self,lanzetti2021interplay,balac2019modeling}, and between \glspl{abk:msp} and users~\cite{dandl2019autonomous,lei2018evolutionary}.
While~\cite{kuiteing2017network,bimpikis2019spatial,chremos2022analytical} focus on pricing strategies for \glspl{abk:msp} at the network level,~\cite{yang2019subsidy,chen2016management} leverage game theory to study management and subsidies of fleets of shared electric vehicles.
Game-theoretic approaches to regulate congestion leveraging particular policies have been studied in~\cite{swamy2012effectiveness,paccagnan2021optimal,lazar2020optimal}, via optimized routing and pricing in~\cite{krichene2017stackelberg}, and via tolling policies in~\cite{bianco2016game}.
In~\cite{zardini2021game} we propose the first game formulation of a complete mobility system, considering equilibria across interactions between authorities, public transit, \glspl{abk:msp}, and users.
Apart from game-theoretic tools, there has been research on policy-making for future mobility systems, focusing on reducing externalities and producing socially efficient solutions~\cite{fullerton2002can,iwata2016can,zhang2016optimal,chremos2020socially,zardini2022co,chremos2023traveler,zoepf2018economics,ostrovsky2019carpooling}.

\paragraph{Statement of Contribution}
In this paper, we provide a game-theoretic formulation to study interactions in multi-modal mobility systems, whereby citizens have access to several, but limited transportation options (e.g., due to bounds on fleet sizes) and seek individually-optimal decisions (e.g., minimizing travel time and fare). 
We then study the arising game. 
Among others, we prove existence of an equilibrium and show that an equilibrium results from the solution of a convex optimization problem. 
Finally, we perform a case study for the classic scenario of Sioux Falls, South Dakota, USA. 
Beside presenting the nominal behavior of our model, we perform sensitivity analyses to study the impact of parameters such as prices and fleet sizes and show how our game-theoretic model of the mobility systems can be embedded in a larger game, among stakeholders of the mobility ecosystems, allowing us to reason about the strategic interventions in the mobility ecosystem.

%% file: chapters/model.tex
\section{Modeling interactions in multi-modal mobility systems as non-atomic games}

In this section, we present a model for the interactions between citizens traveling in a multi-modal mobility system. 
Our model builds on non-atomic routing games with capacitated edges~\cite{correa2004selfish}.
Proofs are provided in the Appendix.

\subsection{Modeling of the Population}
We consider~$N$ locations in a city, indexed by~$i\in\{1,\ldots,N\}$. 
From each location~$i$, citizens want to travel to a specific location~$j\in\{1,\ldots,N\}\setminus\{i\}$. 
Citizens are divided into~$K$ populations, each representing a given category of citizens (e.g., employees, students, etc.). 
We model each population of citizens as a continuum; accordingly,~$d_{ijk}\in \nonnegreals$ is the \emph{demand} of users (i.e., number of users per unit time) of population~$k$ travelling from location~$i$ to location~$j$, with $d_{iik}=0$ for all $i\in\{1,\ldots,N\}$ and $k\in\{1,\ldots,K\}$.

\subsection{Modeling of the \glspl{abk:msp}}
Citizens have several options to reach their destination. 
They can either use the service of one of the \glspl{abk:msp}, or walk to their destination. 
Specifically, we consider~$M$ different \glspl{abk:msp} providing mobility services. 
Each mobility provider~$m$ has~$\availableveh{i}{m}$ available vehicles at location~$i$ (e.g., due to limited fleet size) and can displace at most~$\displaceveh{j}{m}$ vehicles to location~$j$ (e.g., due to limited parking spots). 
The maximum number of customers on a ride from~$i$ to~$j$ is~$\maxrideveh{i}{j}{m}$.
For notational convenience, we associate the index~$m=0$ to walking.
Moreover, we drop the corresponding capacity constraints (i.e.,~$\displaceveh{i}{0}=\availableveh{j}{0}=\maxrideveh{i}{j}{0}=+\infty$ for all~$i,j\in\{1,\ldots,N\}$), so that walking is always a feasible transportation mode.

\subsection{Cost of Traveling}
To each population, origin-destination pair, and mode of transport, we associate a cost function~$\costride{i}{j}{m}\colon \nonnegreals\to\nonnegreals$, which includes travel time, fare (converted to time via the monetary value of time), and possibly other aspects, such as discomfort and environmental impacts.
Specifically,~$\costride{i}{j}{m}(x)$ is the cost which a citizen travelling from location~$i$ to location~$j$ with mode of transport~$m$ incurs in when there are~$x$ citizens (e.g., belonging to another population) who are also travelling from~$i$ to~$j$ with the same mode.
A classical example is 
\begin{equation}
\label{eq:example_cost}
c_{ijk}^m(x)=t_{ij,\mathrm{nom}}^m(1+\alpha_{ij}^m x) + \frac{p_{ij}^m}{V_{T,k}},
\end{equation}
where~$t_{ij,\mathrm{nom}}^m$ is the nominal travel time,~$\alpha_{ij}^m$ is a positive parameter indicating how much congestion (i.e.,~$x$) affects travel time,~$p_{ij}^m$ is the price of the ticket, and~$V_{T,k}$ is the value of time for population~$k$. 
Another popular example in road networks literature, which is the so-called BPR function which relates the occupancy of a road with its travel time: 
\begin{equation}
\label{eq:example_cost_BPR}
c_{ijk}^m(x)=t_{ij,\mathrm{nom}}^m\left(1+\alpha_{ij}^m \left(\frac{x}{\kappa_{ij}^m}\right)^\beta\right) + \frac{p_{ij}^m}{V_{T,k}},
\end{equation}
where~$\beta$ is a coefficient determining the threshold at which the function will rise significantly,~$\kappa_{ij}^m$ is the nominal capacity of the road, and other parameters are as in \eqref{eq:example_cost}.

\subsection{Equilibrium}
We collect the decisions of the (non-atomic) citizens in the non-negative variables~$\{x_{ijk}^{m}\}$, henceforth referred to as \emph{flows} or \emph{configuration}. 
In particular,~$x_{ijk}^{m}$ indicates the flow of customers of population~$k$ travelling from~$i$ to~$j$ who opted for the transportation mode~$m$ to complete their trip. 
Clearly, we are interested in configurations which satisfy the transportation demand and respect the capacity constraints.
We call such configurations \emph{feasible}.

\begin{definition}[Feasible configuration]
A configuration~$\{x_{ijk}^{m}\}$ is \emph{feasible} if
\begin{enumerate}
\item $x_{ijk}^m$ is non-negative;
\item it serves the transportation demand, i.e.,
\begin{equation*}
    \sum_{m}x_{ijk}^m=d_{ijk};
\end{equation*}
\item it satisfies the constraints, i.e., 
\begin{equation*}
    \sum_{k} x_{ijk}^m\leq C_{ij,\mathrm{r}}^m,\quad 
    \sum_{i, k}x_{ijk}^m\leq C_{j,\mathrm{d}}^m,\quad 
    \sum_{j, k}x_{ijk}^m\leq C_{i,\mathrm{a}}^m; 
\end{equation*}
\end{enumerate}
\end{definition}

Given this definition, we can now define \emph{equilibiria}. 
In this context, we say that a configuration~$\{x_{ijk}^{m}\}$ is a Nash equilibrium if no agent can reduce its travel cost by switching to another transportation mode. 
Feasibility represents the least requirement for an equilibrium. 
The other one is \emph{optimality}, given the decisions of all other players. 
Formally: 
\begin{definition}[Nash Equilibrium]
Let~$\{x_{ijk}^{m}\}$ be a feasible flow. 
We say that~$\{x_{ijk}^m\}$ is a \emph{Nash equilibrium} of the game if for all~$i,j,k,m$ with~$x_{ijk}^{m}>0$ every other mode~$m'\in\{0,\ldots,M\}$ either (i) leads to higher cost, i.e.,
\begin{equation*}
c_{ijk}^m\left(\sum_{k}x_{ijk}^m\right)
    \leq 
c_{ijk}^{m'}\left(\sum_{k}x_{ijk}^{m'}\right),
\end{equation*}
or (ii) it is saturated, i.e., 
\begin{equation*}
    \sum_{k} x_{ijk}^{m'}= C_{ij,\mathrm{r}}^{m'}\text{ or } 
    \sum_{i, k}x_{ijk}^{m'}= C_{j,\mathrm{d}}^{m'}\text{ or } 
    \sum_{j, k}x_{ijk}^{m'}= C_{i,\mathrm{a}}^{m'}.
\end{equation*}
\end{definition}
Intuitively, all modes used at equilibrium must yield minimum cost among all modes which are not saturated; else, an agent would transition to a mode with lower cost.

\subsection{Analysis of the Game}
The main tool for the analysis of the game consists of a reformulation as a convex optimization problem, which stems from a reformulation of the game as a non-atomic routing game with capacitated edges~\cite{correa2004selfish}.
Our reformulation proves \emph{existence} of equilibria and serves as an efficient computational routine for their computation.

To recover our results, we require a few mild assumptions on the cost of travelling: 
\begin{assumption}[Regularity of the cost function]\label{ass:cost}
The cost functions~$c_{ijk}^m\colon \mathbb{R}_{\geq 0}\to\mathbb{R}_{\geq 0}$ can be decomposed (with a slight abuse of notation) as 
\begin{equation*}
    c_{ijk}^m(x)
    =
    c_{ijk}^m + c_{ij}^m(x),
\end{equation*}
where~$c_{ijk}^m\in\mathbb{R}_{\geq 0}$ and~$c_{ij}^m\colon \mathbb{R}_{\geq 0}\to\mathbb{R}_{\geq 0}$ are non-decreasing and lower semi-continuous.
\end{assumption}

In words,~\cref{ass:cost} predicates that the term of the cost function coupling the agent is population-independent (i.e., it does not depend on~$k$). 
This assumption is readily satisfied when an agent seeks to minimize travel time or the ``sum'' of travel time and ticket price (see examples in \eqref{eq:example_cost} and \eqref{eq:example_cost_BPR}). 
Then, our result takes the following form: 

\begin{theorem}[Equilibria of the game]\label{thm:equilibria}
Let~\cref{ass:cost} hold. 
Let~$\{x_{ijk}^{m}\}$ be a feasible configuration resulting from the convex optimization problem
\begin{subequations}\label{eq:optimization}
\begin{align}
    \min_{x_{ijk}^m}\: &
    \sum_{i,j,k,m} c_{ijk}^m x_{ijk}^m
    +
    \sum_{i,j,k} \int_0^{\sum_k x_{ijk}^m}c_{ij}^m(z)\mathrm{d}z,
    \\
    \mathrm{s.t. }\:
    &\sum_{m}x_{ijk}^m=d_{ijk},
    \\
    &\sum_{j, k}x_{ijk}^m\leq C_{i,\mathrm{a}}^m,
    \\
    &\sum_{i, k}x_{ijk}^m\leq C_{j,\mathrm{d}}^m,
    \\
    &\sum_{k}x_{ijk}^m\leq C_{ij,\mathrm{r}}^m,
    \\
    &x_{ijk}^m\geq 0.
\end{align}
\end{subequations}
Then,~$\{x_{ij}^{k,m}\}$ is an equilibrium. 
In particular, an equilibrium \emph{always} exists.
\end{theorem}

\begin{remark}[Convexity]
    The optimization problem~\eqref{eq:optimization} is indeed convex. 
    All constraints are clearly linear in the decision variables. 
    To show convexity of the objective function observe that (i)~$\int_0^x c_{ij}^m(z)\mathrm{d}z$ is convex since~$c_{ij}^m$ is non-decreasing~$\mathbb{R}_{\geq 0}$, and so its integral is convex on~$\mathbb{R}_{\geq 0}$, (ii) the composition of a convex function and a linear function (namely,~$(x_{ij1}^m,\ldots,x_{ijK}^m)\mapsto\sum_{k}x_{ijk}^m)$) is convex; (iii) the sum of convex functions is convex. 
    In particular, the optimization problem has~$N(N-1)MK$ decision variables and~$\mathcal{O}(N(N-1)MK)$ constraints (since~$x_{iik}^m=0$ for all~$i\in\{1,\ldots,N\}$).
\end{remark}

In a nutshell, \cref{thm:equilibria} states that an equilibrium of the game results from a minimizer of a convex optimization problem, which can be efficiently computed by off-the-shelf solvers even in large instances.  
The ``converse'' of~\cref{thm:equilibria} does not necessarily hold: There might exist equilibria which are not minimizers of~\eqref{eq:optimization}, as the next example shows. 

\begin{example}\label{ex:converse}
Consider the network reported in \cref{fig:example_theorem}.
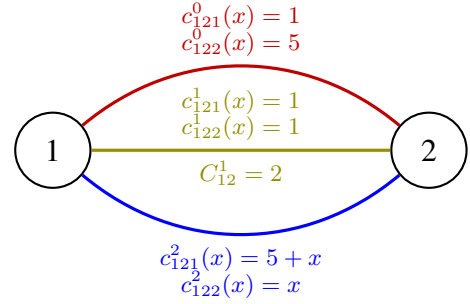
\begin{figure}[tb]
\begin{center}
\begin{tikzpicture}
    \node[circle,thick,draw,minimum size=1cm] at (0,0) (one) {\large 1}; 
    \node[circle,thick,draw,minimum size=1cm] at (5,0) (two) {\large 2};

    \draw[-,very thick,red] (one) to[bend left=40] node[pos=0.5,above,align=center,font=\small] {$c_{121}^0(x)=1$ \\ $c_{122}^0(x)=5$} (two);
    \draw[-,very thick,olive] (one) -- node[pos=0.5,above,align=center,font=\small] {$c_{121}^1(x)=1$ \\ $c_{122}^1(x)=1$} node[pos=0.5,below,align=center,font=\small] {$C_{12}^1=2$} (two);
    \draw[-,very thick,blue] (one) to[bend right=40] node[pos=0.5,below,align=center,font=\small]  {$c_{121}^2(x)=5+x$ \\ $c_{122}^2(x)=x$} (two);
\end{tikzpicture}
\end{center}
\caption{Exemplary network, used to show that the converse of \cref{thm:equilibria} does not necessarily hold. 
Here, all capacity constraints are dropped, except for the green edge, which has capacity 2.}
\label{fig:example_theorem}
\end{figure}
Suppose that there are two populations (i.e.,~$K=2$), the demand from node 1 to node 2 is 2 for both populations (i.e.,~$d_{121}=d_{122}=2$), and that the green edge is capacited to 2 (whereas all other capacities are infinity). 
It is easy to see that, at any equilibrium, the red edge is never used by population 2 (since it leads to a delay of 5, which is always worse than the blue edge) and that the blue edge is never used by population 1 (since it leads to a delay of at least 5, which is always worse than the red edge). 
Let~$x_k\in[0,2]$ be the equilibrium flow through the green edge of population~$k$. 
Because of the edge capacity, it necessary holds~$x_1+x_2\leq 2$. 
For this simple system we can find equilibria by inspection. 
In particular, we distinguish two cases: 
\begin{enumerate}
    \item $x_2=1$ and so~$x_1\in[0,1]$. This configuration leads to a delay of 1 for all agents of both populations. Since no agent can achieve a lower cost by switching to an alternative mode, this is indeed an equilibrium.
    \item $x_1\in(1,2]$ and~$x_2\in[0,1)$ so that~$x_1+x_2=2$. 
    In this case, the delay for all agents of population 1 is 1, whereas the agents of population 2 traversing the green edge incur a delay of 1, and the ones traversing the blue edge in a delay of~$2-x_2>1$. 
    Still, the capacity of the green edge hinders agents on the blue edge to transfer to the green edge, so no agent can obtain a lower cost by switching mode of transport. 
    Thus, this is also an equilibrium. 
\end{enumerate}
Let us now instead deploy~\cref{thm:equilibria} and solve the optimization problem~\eqref{eq:optimization}.
As above, the red edge is never used by population 2, and the blue edge is never used by population 2. So the optimization problem~\eqref{eq:optimization} reduces to
\begin{align*}
    \min_{x_1, x_2}\: &
    (2-x_1) + x_1 + x_2 + \frac{1}{2}(2-x_2)^2 
    \\
    \mathrm{s.t. }\:
    &x_1+x_2\leq 2,
    \\
    &x_1,x_2\geq 0.
\end{align*}
The optimization problem admits the optimal solution~$x_2^\ast=1$ and~$x_1^\ast\in [0,1]$, which yields an objective value of~$7/2$. 
This is precisely the equilibrium in 1).
In particular, the equilibrium in 2) leads to an objective value of~$2+x_2+\frac{1}{2}(2-x_2)^2>7/2$ for all~$x_2<1$. Thus, the equilibrium in 2) does not result from the optimization problem~\eqref{eq:optimization}.
\end{example}

\begin{remark}
In the literature, equilibria resulting from optimization problems akin to~\eqref{eq:optimization} are referred to as Beckmann-McGuire-Whinston (BMW) equilibria~\cite{correa2004selfish}. They enjoy many properties which make them particularly realistic (e.g., for our case studies). We mention two. 
First, a BMW equilibrium is an uncapacitated equilibrium for the game where the shadow prices (i.e., dual multiplier of the capacity constraints) is included as an additional cost. 
Second, consider a surrogate game where capacity constraints appear as barrier terms in the latency functions. Then, as the barrier parameter goes to zero, citizens behave as at a BMW equilibrium.
\end{remark}

To obtain an if and only if statement in~\cref{thm:equilibria}, we need to drop the capacity constraints.
In this case, we can decouple the game in~$N(N-1)$ independent games and deploy standard results in non-atomic routing games~\cite{roughgarden2007routing}:

\begin{corollary}[Equilibria of non-capacitated networks]\label{cor:uncapacitated}
Let~\cref{ass:cost} hold and assume there is no capacity; i.e.,~$C_{i,\mathrm{d}}=C_{i,\mathrm{a}}=C_{ij,\mathrm{r}}^m=+\infty$. 
Then, a feasible configuration~$\{x_{ijk}^m\}$ is an equilibrium if and only if for all~$i,j\in\{1,\ldots,N\}$ $\{x_{ijk}^m\}_{km}$ lies in the set of minimizers of the following optimization problem (parameterized in~$i$ and~$j$):
\begin{subequations}\label{eq:optimization uncapacity}
\begin{align}
    \min_{x_{ijk}^m}\: &
    \sum_{k,m} c_{ijk}^m x_{ijk}^m
    +
    \sum_{m} \int_0^{\sum_k x_{ijk}^m}c_{ij}^m(z)\mathrm{d}z
    \label{eq:cor:no_capacity:objective}
    \\
    \mathrm{s.t. }\:
    &\sum_{m}x_{ijk}^m=d_{ijk},
    \\
    &x_{ijk}^m\geq 0.
\end{align}
\end{subequations}
\end{corollary}

The optimization problem~\eqref{eq:optimization uncapacity} provides us with a full characterization of equilibria when all capacity constraints are dropped. 
It might be tempting to claim uniqueness of the equilibria when the costs~$c_{ij}^m$ are strictly increasing and, thus, its integral the strictly convex. 
However, because of the term~$\sum_kx_{ijk}^m$, the objective function~\eqref{eq:cor:no_capacity:objective} only remains convex and so equilibria are generally not unique (see the next example). 

\begin{example}[Non-uniqueness]
Consider the network reported in \cref{fig:example_uniqueness}, where all edges have strictly monotone costs and are uncapacitated. 
\begin{figure}[tb]
\begin{center}
\begin{tikzpicture}
    \node[circle,thick,draw,minimum size=1cm] at (0,0) (one) {\large 1}; 
    \node[circle,thick,draw,minimum size=1cm] at (5,0) (two) {\large 2}; 
    
    \draw[-,very thick,red] (one) to[bend left=20] node[pos=0.5,above,align=center,font=\small] {$c_{121}^0(x)=x$ \\ $c_{122}^0(x)=x$} (two);
    \draw[-,very thick,blue] (one) to[bend right=20] node[pos=0.5,below,align=center,font=\small]  {$c_{121}^1(x)=x$ \\ $c_{122}^1(x)=x$} (two);
\end{tikzpicture}
\end{center}
\caption{Exemplary network showing that equilibria are generally not unique, even if the cost functions are strictly monotone.}
\label{fig:example_uniqueness}
\end{figure}
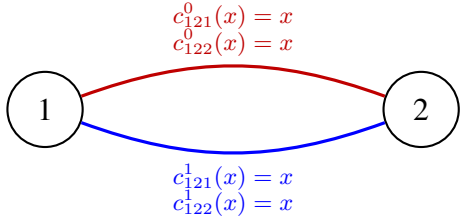
Suppose, again, that there are two populations with demand 2 from node 1 to node 2 (i.e.,~$d_{121}=d_{122}=2$). 
With~$x_k\in[0,2]$ being the flow corresponding to population~$k$ on the red edge, it is readily verified that the optimization problem~\eqref{eq:optimization uncapacity} (for~$i=1$ and~$j=2$) reduces to
\begin{align*}
    \min_{x_1,x_2}\: &
    \frac{1}{2}(x_1+x_2)^2+\frac{1}{2}((2-x_1)+(2-x_2))^2
    \\
    \mathrm{s.t. }\:
    &x_1, x_2\geq 0.
\end{align*}
It is readily checked that any~$(x_1,x_2)\in\{(\alpha,2-\alpha)\colon \alpha\in[0,2]\}$ is optimal and yields a cost of~$2$ for all agents of both populations. By~\cref{cor:uncapacitated}, all these configurations are equilibria of the game. 
In particular, the game possesses infinitely many equilibria. 
\end{example}

\subsection{Discussion}
A few comments are in order. 
First, we abstract away from the topology of the road and public transport network. 
This way, we are implicitly not allowing users to combine different transportation modes to reach their destination (i.e., complete part of the trip by bus and part by bike). 
We defer this analysis to future research.
Second, we assume that different transportation modes do not interfere. 
Nevertheless, these assumptions can be readily relaxed via a slight modification of the travel costs~\cite{roughgarden2005selfish}.
Third, if the capacity constraints~$C_{i,\mathrm{a}}^m$ and~$C_{j,\mathrm{d}}^m$ are dropped, our game can be decomposed as~$N(N-1)$ games, one for each origin-destination pair, which can be solved independently (as done, for instance, in~\cref{cor:uncapacitated} for the special case~$C_{ij}^m=+\infty$). 
Formally, this translates into an optimization problem which can be decomposed into parallel independent subproblems.

%
\begin{comment}
\subsection{An Illustrative Example}

As as example, consider the setting shown in~\cref{fig:example}.

\begin{figure}[!h]
\begin{center}
\begin{tikzpicture}
    \node[circle,thick,draw,minimum size=1cm] at (0,0) (one) {\large 1}; 
    \node[circle,thick,draw,minimum size=1cm] at (4,0) (two) {\large 2}; 
    \node[circle,thick,draw,minimum size=1cm] at (2,-2.83) (three) {\large 3}; 
    
    \draw[-,very thick,red] (one) to[bend left=20] (three);
    \draw[-,very thick,yellow] (one) -- (three);
    \draw[-,very thick,blue] (one) to[bend right=20] (three);
    
    \draw[-,very thick,red] (one) to[bend left=20] (two);
    \draw[-,very thick,yellow] (one) -- (two);
    \draw[-,very thick,blue] (one) to[bend right=20] (two);
    
    \draw[-,very thick,red] (two) to[bend left=20] (three);
    \draw[-,very thick,yellow] (two) -- (three);
    \draw[-,very thick,blue] (two) to[bend right=20] (three);
\end{tikzpicture}
\end{center}
\caption{Network for the illustrative example, with three transportation modes (including walking).}
\label{fig:example}
\end{figure}

For simplicity, we consider a single, homogeneous population (i.e., $K=1$), two transportation modes (i.e., $M=2$) as well as the option of walking, and unit transportation demand for each origin-destination pair. 
The cost functions are\ldots

Parameters:
\begin{itemize}
\item Fleet sizes per zone, per mode;
\item Prices per ride (PT), per km (others);
\item Taxes on different services;
\item The role of $\alpha$
\item 
\end{itemize}
\end{comment}

%% file: chapters/results.tex
\section{Case studies}
\label{sec:results}
We consider the classic case study of Sioux Falls, South Dakota, USA.
Here, locations correspond to the 24 different nodes of the well-known network, and each pair of locations is characterized by particular demand patterns, which can be inferred from the literature~\cite{TNTP23}.
We consider a time window of three hours during the weekly morning peak.
For the sake of the case study, we randomly split the travelers into three populations, consisting of students, business travelers, and leisure travelers.
We consider four modes of transportation: public buses, \gls{abk:amod}, shared bikes, and walking. 
To each origin-destination pair, mode of transportation, and population we associate the cost function
\begin{equation*}
    c_{ijk}^m(x)=\frac{p_{ij}^m}{V_{T,k}} + t_{ij}^m(x),
\end{equation*}
where~$p_{ij}^m$ is the price of the ticket for transportation mode~$m$,~$V_{T,k}$ is the value of time of population~$k$, and~$t_{ij}^m(x)$ is the travel time of transportation mode~$m$ between locations~$i$ and~$j$. 
We detail the choices in the following.
\paragraph*{Price and value of time}
For the public transportation system,~$p_{ij}^m$ represents the price of a ride (and it is independent of~$i$ and~$j$). 
We consider a nominal ride cost of \SI{2.5}{\usd}.
For \gls{abk:amod} and shared bike services, the price, instead, depends on the length of the trip. 
We model it via a distance-based of \SI{2.5}{\usd\per\kilo\meter} for \gls{abk:amod} and for \SI{0.4}{\usd\per\kilo\meter} for bikes. 
Walking is clearly considered free of charge. 
Finally, the value of time~$V_{T,k}$ only varies across populations. 
In line with existing studies~\cite{lanzetti2021interplay}, we consider values of time of \SI{15}{\usd\per\hour} for students, \SI{35}{\usd\per\hour} for business travelers, and \SI{7}{\usd\per\hour} for leisure travelers. 

\paragraph*{Travel time}
For buses, shared bikes, and walking we assume that travel time is independent of the total load between~$i$ and~$j$ (i.e.,~$\sum_{k}x_{ijk}^m$) and therefore directly consider~$t_{ij}^m(x) = t_{ij}^\mathrm{nom}$.
We infer the nominal travel time~$t_{ij}^\mathrm{nom}$ from the distance between~$i$ and~$j$ assuming standard average speeds for all modes.
For \gls{abk:amod}, instead, we consider road congestion, which we model via the so-called BPR function (cf.~\cref{eq:example_cost_BPR}), so that 
\begin{equation*}
    t_{ij}^m(x) = t_{ij}^\mathrm{nom}\left(1+\alpha\left(\frac{x}{\kappa}\right)^\beta\right),
\end{equation*}
with~$\alpha=0.15$,~$\beta=4$, and~$\kappa=250$.

We impose capacity constraints based on vehicle availability at each station, over the considered time window. 
For \gls{abk:amod} and shared bikes, we assume a fleet of 1,200 vehicles, which we uniformly split over the available locations. 
For public transport, we assume 120 buses per location and that each bus can host up to 50 travelers. 
Clearly, walking is uncapacitated. 
For simplicity, we do not impose additional constraints on edge and arrival rate. 
In particular,~$C_{ij,\mathrm{r}}^m=+\infty$ (i.e., no maximum number of vehicles of a given transportation mode on edge~$(i,j)$) and~$C_{j,\mathrm{d}}^m=+\infty$ (i.e., no maximum number of vehicle of transportation mode can arrive at~$j$).

We find an equilibrium of the game by solving the convex optimization problem~\eqref{eq:optimization}, as predicated by~\cref{thm:equilibria}.
To avoid non-reproducibility of our results, related to the non-uniqueness of solutions for the optimization problems~\eqref{eq:optimization}, we use a quadratic regularization with weight 0.001. 
The optimization problem is formulated in CVXPY~\cite{diamond2016cvxpy} and solved with ECOS~\cite{domahidi2013ecos}. 
On commodity hardware (MacBook Pro, 2.3 GHz Quad-Core Intel Core i5, RAM 16GB) the solving time amounts to approximately \SI{1}{\second}.

Given the presented model, we illustrate the following three case studies, which highlight the capabilities of the framework.
First, we inspect its nominal behavior.
Second, we use our model to reason about the impact of a strategic relocation of vehicles. 
Third, we integrate our model in a game-theoretic framework which models the interactions between mobility stakeholders, all the way from municipalities to citizens, through mobility providers.

\subsection{Nominal Case}
To start, we illustrate the nominal behavior of the model. 
At equilibrium, 49.9\% of the population travels via public buses, 17.3\% chooses \gls{abk:amod} services, 30.6\% opts for bike sharing, and 2.2\% walks. 
In general, business travelers, who have a higher value of time, prefer to reach their destination using the (usually) fastest transportation mode, namely \gls{abk:amod}; indeed, 46.8\% of business travelers travel by car. 
Conversely, 62.8\% of leisure travelers, the population with the lowest value of time, travel by bike, which is the cheapest transportation options for most origin-destination pairs.
The average travel cost (including fare and monetary value of time) for a citizen is~\SI{6.88}{\usd}. 
On average, however, business travelers pay \SI{11.20}{\usd} for their trip, leisure travelers only pay~\SI{3.48}{\usd}, and students pay~\SI{6.14}{\usd}, in line with their value of time. 

\subsection{The Impact of Strategic Relocation of Vehicles}
\begin{figure}[tb]
    \centering
    \includegraphics[width=7cm]{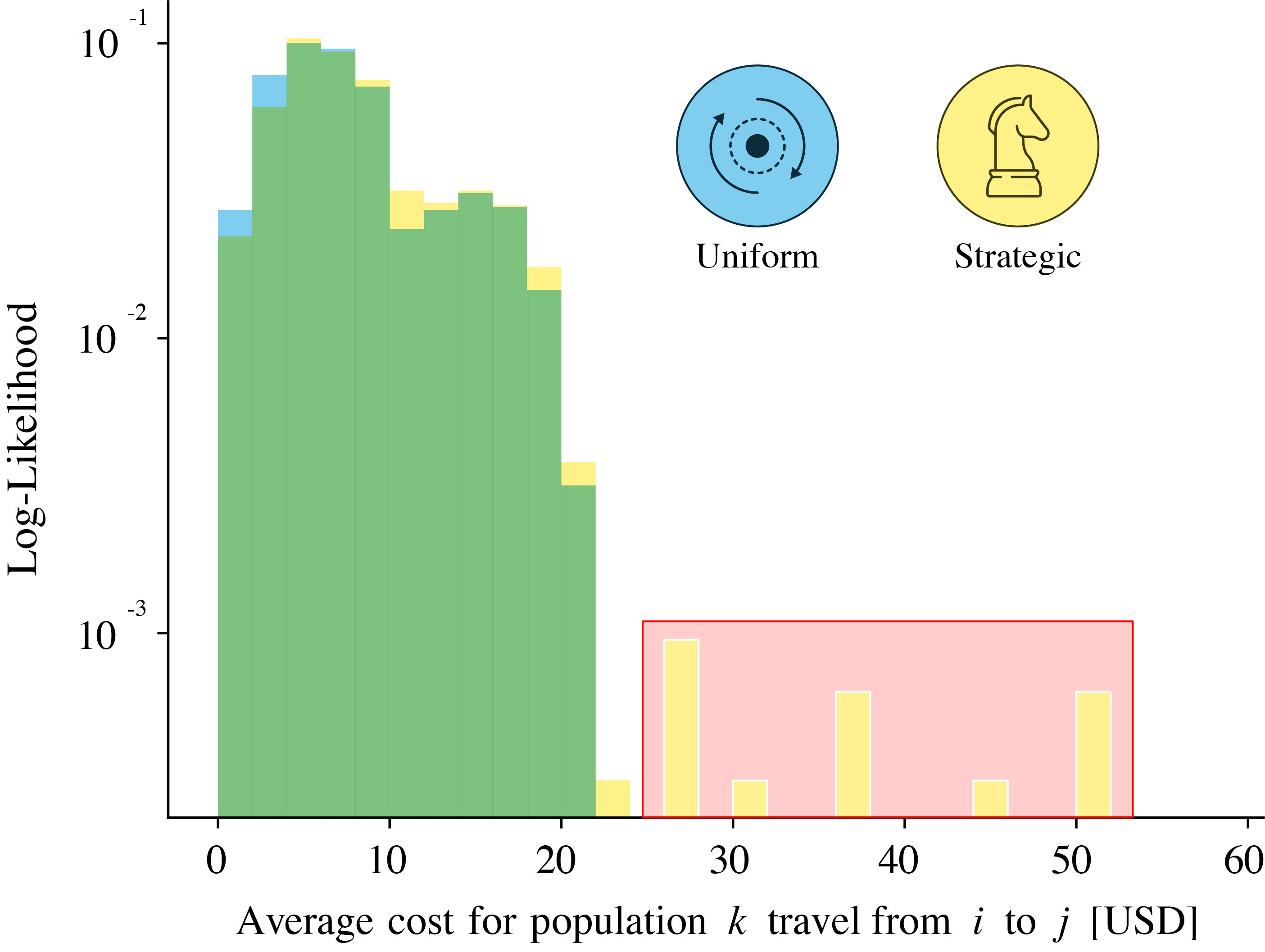}
    \caption{Log-likelihood of the travel cost for population~$k$ travelling from~$i$ to~$j$; i.e., the histogram results from~$N(N-1)K$ populations ($N(N-1)$ origin-destination pairs for each of the~$K$ populations). 
    With a strategic allocation of vehicles, \gls{abk:amod} and bike operator can increase their revenues. This way, however, a few populations face much higher transportation costs, as the tail of the orange distribution shows.}
    \label{fig:relocation}
\end{figure}
Our second case study pertains to the impact of relocation strategies for \gls{abk:amod} and shared bikes, which effectively determines the availability of vehicles~$C_{\mathrm{a},i}$ at each location~$i$. 
For this case study, we assume that at each location~$i$ the available buses can serve up to 70\% of the total transportation demand starting their trip at location~$i$; for \gls{abk:amod} and bikes, instead, we consider a constant fleet size (1,200 vehicles) but with two different spacial distributions of vehicles: (i) Vehicles are relocated uniformly across all locations and so the availability of vehicles is~$N_\mathrm{fleet}/N$ at each location, as in our nominal case; (ii) vehicles are relocated uniformly across the 66\% most densely populated locations (whereby we estimate the population density of a location via the number of trips starting at that location), so~$N_\mathrm{fleet}/(0.66N)$ vehicles are available at 66\% of the locations. 

Notably, with strategic relocation of vehicles, the revenue of the \gls{abk:amod} operator increases by 5\% and the revenue of the bike operator by 9\%, given the more efficient placement of the vehicles.
However, while the average travel cost across the whole population is nearly unchanged, several populations and origin-destination pairs face significantly larger travel costs (see~\cref{fig:relocation}), mainly because they are left only with inconvenient transportation options. 
Conversely, there is a slight increase in citizens with very low average travel costs, since in densely populated areas there are more vehicles and citizens have access to more transportation options. 
Overall, such observations show that, while macroscopic metrics such as profits and average travel costs over the whole population appear unaltered or even improving, there might be undesired microscopic underlying consequences. 
This kind of data provides policy makers with actionable information to design regulations preventing such consequences.

\subsection{A Game between Mobility Stakeholders}
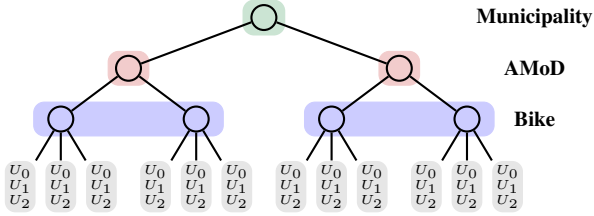
\begin{figure}[t]
    \centering
    \input{img/multistage_game}
    \caption{Sequential interactions of the game in the case of a municipality with two available actions and two \glspl{abk:msp} having two and three available actions, respectively. 
    First, the municipality chooses its action. Then, \gls{abk:amod} and bike operators simultaneously decide on their action. 
    The payoff of all stakeholders follows from a low-level simulator of the mobility system. The boxes depict the so-called information sets: \gls{abk:amod} knows the action of the municipality, but does not know the action of the bike operator.}
    \label{fig:multistage_game}
\end{figure}
In our third case study, we incorporate our model into a game-theoretic framework to study interactions between mobility stakeholders, presented in~\cite{zardini2021game}. 
The interactions are modeled as a two-stage game between mobility stakeholders (see \cref{fig:multistage_game}).
In the first stage, the municipality announces its strategy (e.g., public transport fares); in the second stage, \glspl{abk:msp} simultaneously decide on their actions (e.g., design of their fleet and prices). 
All decisions are then fed to a low-level model of the mobility system, which generates payoffs for all stakeholders.

In this paper, we use our multi-modal model of the mobility system as a low-level simulator. 
We believe that this model provides an attractive trade-off between accuracy and tractability: Analytical models (e.g., see \cite[Section 2D]{zardini2021game}) enable an elegant theoretical analysis but are usually overly simplifying; high-fidelity simulators, instead, are much harder to calibrate and usually require considerable computational time.

We consider the following actions for the municipality, the \gls{abk:amod} operator, and the bike operator:
\begin{center}
    \begin{tabular}{ccc}
        \toprule 
         Municipality & Price per ride [\SI{}{USD}]  &  0,0.5,1.0,\ldots,4.0\\
         AMoD & Price [\SI{}{USD/km}]&0,0.5,1.0,\ldots,4.0 \\
         Shared Bikes & Price [\SI{}{USD/km}]&0,0.4,0.8,1.2\\
         \bottomrule 
    \end{tabular}
\end{center}

For the municipality, the payoff is given by a combination of average travel cost for citizens (comprising fare and travel time), CO\textsubscript{2} emissions (resulting from vehicles' operation, production, etc.), and the revenue resulting from the tickets, respectively weighted with the non-negative parameters~$\rho_\mathrm{cost},\rho_\mathrm{CO\textsubscript{2}}, \rho_\mathrm{revenue}\geq 0$. 
For instance, if~$\rho_\mathrm{cost}>0$ and~$\rho_\mathrm{CO\textsubscript{2}}=\rho_\mathrm{revenue}=0$, the municipality only seeks to minimize citizens' cost; if~$\rho_\mathrm{CO\textsubscript{2}}>0$ and~$\rho_\mathrm{cost}=\rho_\mathrm{revenue}=0$, instead, the municipality only minimizes CO\textsubscript{2} emissions. 
For emissions, we focus on the ones generated by autonomous vehicles (0.1381 kg/km) and shared bikes (0.0205 kg/km)~\cite{zardini2022co}.
We do not penalize the emissions of the public transportation system, as it runs regardless of the volume of citizens using it.
For the \gls{abk:amod} and bike operators, instead, the payoff readily consists of the revenue resulting from fares paid by the citizens. 

As presented in~\cite{zardini2021game}, we find the equilibrium of the sequential (Stackelberg) game of~\cref{fig:multistage_game} via backward induction. 
In turn, this amounts to finding (pure) Nash Equilibria for all possible strategies of the municipality. 
\begin{figure*}[t]
    \includegraphics[width=\linewidth]{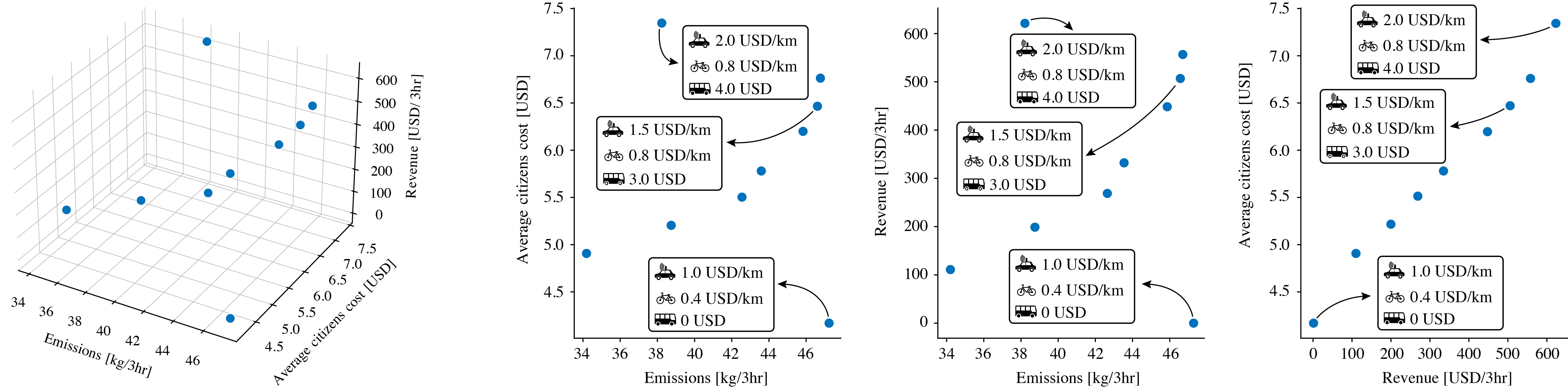}
    \caption{Plot of all Nash Equilibria of the game between \gls{abk:amod} and bike operators w.r.t. the three metrics of the municipality (average travel cost, emissions, and revenue) (left), together with its projection on each pair of axes, with highlighted details for selected equilibria. 
    Among others, these plots highlight that the actions of the municipality are incomparable (i.e., no action is simultaneously best on all metrics) and the optimal action depends on the specific choice of the weight parameters.}
    \label{fig:pareto}
\end{figure*}
In~\cref{fig:pareto}, we plot the outcome of the game between \gls{abk:amod} and bikes for various strategies of the municipality. 
Some configurations are dominated, others are incomparable (i.e., none is simultaneously best on all metrics), and the best one depends on the parameters~$\rho_\mathrm{revenue}, \rho_\mathrm{cost}, \rho_\mathrm{CO\textsubscript{2}}$.
Remarkably, free public transport does not result in the lowest average citizens' costs and emissions. 
In particular, the average citizens' costs increase monotonically with increasing public transport prices; emissions, instead, behave differently. 
Low public transport prices force both the \gls{abk:amod} and shared bike operators to also lower their prices, which makes their service attractive for many citizens. 
With higher prices, instead, the \gls{abk:amod} and bike operators will raise their prices, making their transportation service only attractive to a smaller portion of the population willing to pay a premium for the more time-efficient service. 
This, however, correlates with fewer vehicles on the road and, thus, lower emissions.

\begin{comment}
\subsection{Changing parameters}
The operators of busses, \gls{abk:amod}, and shared bikes services, can choose their respective fleet sizes, and their allocation in the different zones.
Furthermore, they can choose the pricing strategy for their service.

Fixing the choice of \ldots, one can compute the equilibrium

\subsection{Free Public Transport}
\ldots

\begin{itemize}
    \item Price vs. modal split
    \item Emissions
\end{itemize}

\begin{figure}
    \centering
    
    \caption{Caption}
    \label{fig:my_label}
\end{figure}

\begin{figure}[h]
    \centering
    
    \caption{Mode Split as a function of $\alpha$}
    \label{fig:my_label}
\end{figure}
\end{comment}

%% file: img/multistage_game.tex
\begin{tikzpicture}[scale=0.9]
        \draw[rounded corners,fill=blue!20!white,color=blue!20!white] (-3.4, -1.75) rectangle ++(2.8,0.5) {};
        \draw[rounded corners,fill=blue!20!white,color=blue!20!white] (0.6, -1.75) rectangle ++(2.8,0.5) {};
        \draw[rounded corners,fill=red!20!white,color=red!20!white] (1.7, -1) rectangle ++(0.6,0.5) {};
        \draw[rounded corners,fill=red!20!white,color=red!20!white] (-1.7, -1) rectangle ++(-0.6,0.5) {};
        
        \draw[rounded corners,thick,color=ForestGreen!20!white,fill=ForestGreen!20!white] (-0.3, -0.25) rectangle ++(0.6,0.5) {};
        
        \node[circle,draw,thick] at (0,0) (mun) {};
        \node[circle,font=\footnotesize] at (4,0)  {\textbf{Municipality}};
        
        \node[circle,draw,thick] at (-2,-0.75) (msp11) {}; 
        \node[circle,draw,thick] at (2,-0.75) (msp12) {}; 
        \node[circle,font=\footnotesize] at (4,-0.75)  {\textbf{AMoD}};
        
        \node[circle,draw,thick] at (-3,-1.5) (msp21) {};
        \node[circle,draw,thick] at (-1,-1.5) (msp22) {}; 
        \node[circle,draw,thick] at (1,-1.5) (msp23) {}; 
        \node[circle,draw,thick] at (3,-1.5) (msp24) {}; 
        \node[circle,font=\footnotesize] at (4,-1.5)  {\textbf{Bike}};
        
        \tikzstyle{myrectangle} = [rectangle,font=\tiny,align=center,rounded corners,fill=black!10!white,inner sep=1.5pt]
        \node[myrectangle] at (-3.6,-2.5) (o11) {$U_0$ \\ $U_1$ \\ $U_2$}; 
        \node[myrectangle] at (-3.0,-2.5) (o12) {$U_0$ \\ $U_1$ \\ $U_2$}; 
        \node[myrectangle] at (-2.4,-2.5) (o13) {$U_0$ \\ $U_1$ \\ $U_2$}; 
        \node[myrectangle] at (-1.6,-2.5) (o21) {$U_0$ \\ $U_1$ \\ $U_2$}; 
        \node[myrectangle] at (-1.0,-2.5) (o22) {$U_0$ \\ $U_1$ \\ $U_2$}; 
        \node[myrectangle] at (-0.4,-2.5) (o23) {$U_0$ \\ $U_1$ \\ $U_2$}; 
        \node[myrectangle] at (1.6,-2.5) (o31) {$U_0$ \\ $U_1$ \\ $U_2$}; 
        \node[myrectangle] at (1.0,-2.5) (o32) {$U_0$ \\ $U_1$ \\ $U_2$}; 
        \node[myrectangle] at (0.4,-2.5) (o33) {$U_0$ \\ $U_1$ \\ $U_2$}; 
        \node[myrectangle] at (3.6,-2.5) (o41) {$U_0$ \\ $U_1$ \\ $U_2$}; 
        \node[myrectangle] at (3.0,-2.5) (o42) {$U_0$ \\ $U_1$ \\ $U_2$}; 
        \node[myrectangle] at (2.4,-2.5) (o43) {$U_0$ \\ $U_1$ \\ $U_2$}; 
        
        \draw[thick,-] (mun) -- (msp11); 
        \draw[thick,-] (mun) -- (msp12); 
        
        \draw[thick,-] (msp11) -- (msp21); 
        \draw[thick,-] (msp11) -- (msp22); 
        \draw[thick,-] (msp12) -- (msp23); 
        \draw[thick,-] (msp12) -- (msp24); 
        
        \draw[thick,-] (msp21) -- (o11); 
        \draw[thick,-] (msp21) -- (o12); 
        \draw[thick,-] (msp21) -- (o13);
        \draw[thick,-] (msp22) -- (o21); 
        \draw[thick,-] (msp22) -- (o22); 
        \draw[thick,-] (msp22) -- (o23); 
        \draw[thick,-] (msp23) -- (o31); 
        \draw[thick,-] (msp23) -- (o32); 
        \draw[thick,-] (msp23) -- (o33); 
        \draw[thick,-] (msp24) -- (o41); 
        \draw[thick,-] (msp24) -- (o42); 
        \draw[thick,-] (msp24) -- (o43); 
\end{tikzpicture}

%% file: chapters/conclusion.tex
\section{Conclusion}\label{sec:conclusion}
We proposed a game-theoretic framework to study interactions in multi-modal mobility systems.
Our framework builds on non-atomic routing games with capacitated edges and models the behavior of citizens in the presence of various transportation options. 
We study the described game and prove the existence of an equilibrium, which can be found efficiently by solving a convex optimization problem.
We instantiate our framework in the classic case study of Sioux Falls, USA, and show its capabilities to reason about interventions in mobility systems.

This work unlocks exciting venues for future research. 
First, we intend to implement our framework across different categories of basic models for the mobility system. 
This involves precisely defining the equilibria and exploring algorithms to calculate them efficiently. 
Second, we aim to utilize our framework to model and analyze interactions occurring at various time scales (e.g., daily vs. weekly, vs. yearly). 
Lastly, we strive to apply our methodology to examine other scenarios that share similar structures, such as settings in energy and data markets.

%% file: chapters/appendix.tex
\input{chapters/proofs}

%% file: chapters/proofs.tex
\section{Proofs}
\label{sec:appendix}

\begin{proof}[Proof of~\cref{thm:equilibria}]
The proof builds on a reformulation of the game as a standard non-atomic routing game with capacitated edges. 
This allows us then to  access the large body of literature studying such games; e.g., see~\cite{correa2004selfish}. 

To ease the presentation, we will use the following notation: Indices~$i,j$ span over the set~$\{1,\ldots,N\}$, the index~$k$ over the set~$\{1,\ldots,K\}$, and the index~$m$ over the set~$\{0,\ldots,M\}$. 
Consider then the non-atomic routing game with capacitated edges defined on the following graph and travel demand: 
\begin{itemize}
    \item For each location~$i$ we introduce one node, denoted by~$i$. 
    This node will serve as the destination node for all agents traveling to~$i$. 
    \item For each location~$i$ we introduce node~$K$ nodes, denoted by~$ik$.
    A node~$ik$ will serve as the starting node for all agents of population~$k$ whose origin is~$i$.
    \item For each node~$ik$ we add~$NM$ nodes, denoted by~$ijkm$, and edges~$(ik,ijkm)$ with no capacity and (constant) cost~$\ell_{(ik,ijkm)}(x)=c_{ijkm}$. 
    Thus, this edge will be traversed by all agents of population~$k$ leaving from station~$i$ and going to station~$j$ with mode~$m$.
    \item For each location~$i$ and each mode~$m$ we add the ``departure nodes''~$i_\mathrm{leave}^m$. 
    We introduce the directed edges~$(ijkm,i_\mathrm{leave}^m)$ with zero cost and capacity~$C_{i,\mathrm{a}}^m$. 
    This edge will be traversed by all agents who are leaving station~$i$ with mode~$m$.
    \item For each location~$i$ and each mode~$m$ we add the ``arrival nodes''~$i_\mathrm{arrival}^m$. 
    We introduce the directed edges~$(i_\mathrm{arrival}^m,i)$ with zero cost and capacity~$C_{i,\mathrm{d}}^m$. 
    This edge will be traversed by all agents who are arriving at station~$i$ with mode~$m$.
    \item For each pair of stations~$i,j$ and each mode~$m$ we introduce edges ($i_\mathrm{leave}^m,j_\mathrm{arrival}^m$) with cost~$\ell_{(i_\mathrm{leave}^m,j_\mathrm{arrival}^m)}(x)=c_{ij}^m(x)$,
    where the cost~$c_{ij}^m$ is as in~\cref{ass:cost} and capacity~$C_{ij,\mathrm{r}}^m$. 
    This edge will be traversed by all agents leaving station~$i$ and going to station~$j$ via mode~$m$.
    \item The travel demand is as follows: For travel demand~$d_{ijk}$ of population category~$k$ between locations~$i$ and~$j$, we add the demand~$d_{ijk}$ from node~$ik$ to node~$j$. 
\end{itemize}
Since this is now a standard non-atomic game with non-decreasing and lower semi-continuous costs and capacitated edges, the flows~$f_e, f_P$ resulting from the optimization problem
\begin{subequations}\label{eq:proof:opt paper}
\begin{align}
    \min_{f_e,f_P} &\sum_{e\in E}\int_0^{f_e} \ell_e(z)\mathrm{d} z
    \\
    \mathrm{s.t. }
    &\sum_{P\ni e} f_P=f_{e},
    \\
    &\sum_{P\in P_{ijk}} f_P=d_{ijk},\label{eq:proof:demand paper}
    \\
    &f_e\leq C_e, \\
    & f_e,  f_P\geq 0.
\end{align}
\end{subequations}
form an equilibrium of the game (in particular, it is a so-called BMW equilibrium)~\cite{correa2004selfish}. 
Here,~$E$ is the set of edges on the graph,~$\ell_e$ and~$C_e$ are the cost and capacity of edge~$e\in E$, and~$P_{ijk}$ is the set of paths available to population~$k$ to go from node~$i$ to node~$j$.

We now instantiate the optimization problem~\eqref{eq:proof:opt paper} for our setting. 
In our case, each origin-destination pair has~$M$ available paths (i.e., one per mode of transport); e.g., population~$k$ going from~$i$ to~$j$ can select from a pool of~$M+1$ paths (i.e., walking and~$M$ \glspl{abk:msp}):
\begin{equation*}
ik \to ijkm \to i_\mathrm{leave}^m \to  j_\mathrm{arrival}^m \to j
\quad \text{for all } m.
\end{equation*}
Therefore, we can capture the flow on each of these~$M+1$ paths via a non-negative decision variable~$x_{ijk}^m$ and therefore replace~$f_P$ by~$x_{ijk}^m$ in~\eqref{eq:proof:opt paper}. 
Thus, the constraint~\eqref{eq:proof:demand paper} reads
\begin{equation*}
    \sum_{m}x_{ijk}^m=d_{ijk}.
\end{equation*}
We can now express the flows~$f_e$ as a function of~$\{x_{ijk}^m\}$.
Specifically, the total flow on the edge~$(ik,ijkm)$ is therefore precisely~$x_{ijk}^m$, the total flow on edge~$(ijkm,i_\mathrm{leave}^m)$ is~$\sum_{jk}x_{ijk}^m$, the total flow on edge~$(i_\mathrm{leave}^m,j_\mathrm{arrival}^m)$ is~$\sum_k x_{ijk}^m$, the total flow on edge~$(j_\mathrm{arrival}^m, j$) is~$\sum_{ik}x_{ijk}^m$.
Moreover, observe that in our case the only edges with non-zero costs are the edges~$(ijkm,i_\mathrm{leave}^m)$ and~$(ik,ijkm)$.
We can now combine these expressions with~\eqref{eq:proof:opt paper} to get 
\begin{subequations}\label{eq:proof:opt}
\begin{align}
    \min_{x_{ijk}^m} &
    \sum_{i,j,k,m} c_{ijk}^m x_{ijk}^m
    +
    \sum_{i,j,m} \int_0^{\sum_k x_{ijk}^m}c_{ij}^m(z)\mathrm{d}z\label{eq:proof:objective}
    \\
    \mathrm{s.t. }
    &\sum_{m}x_{ijk}^m=d_{ijk},\label{eq:proof:demand}
    \\
    &\sum_{i,k}x_{ijk}^m\leq C_{j,\mathrm{d}}^m,\label{eq:proof:capacity1}
    \\
    &\sum_{j,k}x_{ijk}^m\leq C_{i,\mathrm{a}}^m,\label{eq:proof:capacity2}
    \\
    &\sum_kx_{ijk}^m\leq C_{ij,\mathrm{r}}^m,\label{eq:proof:capacity3}
    \\
    &x_{ijk}^m\geq 0,\label{eq:proof:nonnegativity}
\end{align}
\end{subequations}
which is precisely~\eqref{eq:optimization}.
Since the feasible region of~\eqref{eq:proof:opt} results from linear equalities and inequalities, it is a polyhedron. 
Boundedness (and thus compactness) follows readily from demand satisfaction (constraint~\eqref{eq:proof:demand}) and non-negativity (constraint~\eqref{eq:proof:nonnegativity}).
Moreover, the integral of a (measurable) function is (absolutely) continuous, so the objective~\eqref{eq:proof:objective} is continuous. 
Thus, by Weierstrass theorem,~\eqref{eq:proof:opt} admits a minimizer, which effectively establishes existence of an equilibrium.
\end{proof}

\begin{proof}[Proof of~\cref{cor:uncapacitated}]
In the uncapacitated case, we can replicate the construction of the proof of~\cref{thm:equilibria} and note that it leads to an uncapacitated non-atomic network game, for which equilibria are well-known coincide with the minimizers of~\eqref{eq:proof:opt} where constraints \eqref{eq:proof:capacity1}, \eqref{eq:proof:capacity2}, and \eqref{eq:proof:capacity3} are dropped~\cite{correa2004selfish,roughgarden2005selfish}.
To conclude, note that the optimization problem can be decomposed into~$N^2$ independent optimization problems, one for each~$i$ and~$j$. 
\end{proof}